\documentclass[12pt,onecolumn,showpacs,amssymb,aps,nofootinbib,floatfix]{revtex4-1}
\usepackage{epsfig}
\usepackage{color}
\newcommand{\ave}[1]{\left\langle #1 \right\rangle}
\newcommand{\qave}[1]{\left| \left\langle #1 \right\rangle \right|^2}
\newcommand{\di}{{\rm d}}
\def\wT{{\widehat T}}

\newcommand{\eqcomma}{\phantom{AA},\phantom{AA}}

\newcommand{\lnz}{\ln \mathcal{Z}}

\newcommand{\order}[1]{ \mathcal{O} \left( #1 \right) }

\begin{document}
\title{Fluctuating Relativistic hydrodynamics from Crooks theorem}
\affiliation{Universidade Estadual de Campinas - Instituto de Fisica "Gleb Wataghin"\\
Rua Sérgio Buarque de Holanda, 777\\
 CEP 13083-859 - Campinas SP\\
torrieri@ifi.unicamp.br\\
}
\begin{abstract}
We use the Crooks fluctuation theorem \cite{crooks} together with Zubarev hydrodynamics \cite{zubarev} to develop a bottom-up theory of hydrodynamic fluctuations.   We also use thermodynamic uncertainity relations to estimate bottom-up limits to dissipative transport coefficients.
\end{abstract}
\author{Giorgio Torrieri}
\affiliation{IFGW, Unicamp}
\email{torrieri@ifi.unicamp.br}
\maketitle
\section{Hydrodynamics and microscopic fluctuations}
In recent years, the applicability of relativistic hydrodynamics to heavy ion collisions generated a concerted effort to derive hydrodynamics from underlying statistical mechanics \cite{kodama}.    However, this derivation usually happened via transport theory, and a self-consistent inclusion of stochastic terms is still not available.   This is a potentially crucial flaw, given that hydrodynamics seems to apply to systems of $\order{20}$ degrees of freedom \cite{cms}, where thermal and statistical fluctuations cannot be neglected.

While statistical fluctuations in equilibrium are easily understood via partition functions, fluctuations in hydrodynamics are still not well understood.  The earliest construction, in \cite{landau}, combines thermodynamic uncertainity relations and fluctuation-dissipation relations to  gives Gaussian locally valued autocorelations used in the linearized limit
\[\   \left(
\begin{array}{c}
  \ave{\Delta e(x)\Delta e(x')}\\
  \ave{\Delta p(x) \Delta p(x')} \\
  \ave{\Delta T_{ij}(x)\Delta T_{ij}(x')}
\end{array}
\right)
\sim \delta(x-x') \left(
\begin{array}{c}
  T^2 c_V \ave{e}   \\  \left[ T^2 c_p \oplus T \left(\zeta - \frac{2}{3} \eta\right) \right] \ave{p} \\ T \eta \ave{T_{ij}} \end{array}\right)
 +  \order{(\Delta ...)^{n>2} } \]
a significant amount of work has gone on since \cite{landau,csernai,kovtun,gale,stephanov,stephcrit} to correct, extend and apply this approach to relativistic viscous hydrodynamics.

This is however not entirely satisfactory:  For one, these fluctuations are really appropriate for a linear theory while hydrodynamics is of course strongly non-linear.   Functional techniques, together with Lagrangian hydrodynamics, can in principle overcome this difficulty \cite{ryblewski,giorgio,burch,lagrangian}, although it is far from clear that the functional integral is stable and convergent \cite{nicolis,gripaios}; Numerical techniques suggest there are phase transitions \cite{burch}, and the inclusion of microscopic polarization suggests its interactions with vorticity could regularize the instabilities \cite{tinti,gt3,linear}.

This is however still not entirely satisfactory either:   Lagrangian hydrodynamics coarse-grains the fluid at the level of volume elements.    Stochastic fluctuations {\em within the volume element itself} are averaged out.   This makes it doubtful weather the quantum microscopic fluctuations are really amenable to this approach, yet they are certainly non-negligible in collisions with $\order{50}$ particles final state. In water, a cube of a side of $\eta/(sT)$ in natural units would have $\order{10^9}$ molecules, and since \[\ P\left( N \ne \ave{N} \right) \sim \exp \left[ -\frac{\left(N-\ave{N}\right)^2}{\ave{N}}\right]\] we can be reasonably certain that at hydrodynamic scales the probability for a significant deviation from the mean is {\em small}.   Quantitatively this can be recast into the hierarchy of scales necessary for hydrodynamics to be a good effective theory \cite{ryblewski}
\begin{equation}
  \label{hyerarchyscales}
 \underbrace{s^{-1/3} \ll \frac{\eta}{sT}}_{Ratio:\alpha} \eqcomma \underbrace{\frac{\eta}{sT} \ll \frac{1}{\partial_\mu u_\nu}}_{Ratio:K} 
\end{equation}
the expansion in Knudsen number K \cite{kodama} is an expansion around the last two quantities, and including fluctuations fully is equivalent to building an effective theory around the first inequality, $\alpha \ll 1$ (note that in the planar limit it is suppressed by the number of colors to the $1/3$).
Experimental data, however, seems to suggest \cite{cms} that systems with 50 particles are in some sense ``just as collective'' as those of 1000.
For such small systems, fluctuations {\em cannot} be small and, given space gradients, the first inequality in Eq. \ref{hyerarchyscales} {\em cannot} hold even if $\eta/s \rightarrow 0$. Even if particles are somehow ``born in equilibrium'' \cite{becborn} ``at every point in space'' \cite{hartnoll}, equilibrium just means all microstates are equally likely, and most likely state is not a certainty.

Within a fully quantum picture, the energy-momentum tensor $T_{\mu \nu} \rightarrow \hat{T}_{\mu \nu}$ becomes an operator.  Any ``local equilibrium decomposition'' of it (with an equilibrium $T_0^{\mu\nu}$ and a dissipative part $\Pi^{\mu \nu}$) would have to be done at the operator level.  I.e. we would have to find a mieaningful way to define
    \begin{equation}
       \label{piTdef}
   \hat{T}^{\mu \nu} = \hat{T}_0^{\mu \nu}+    \hat{\Pi}^{\mu \nu}
\end{equation}
    must be operator-valued.  Later, and in the appendix, we shall define precisely what this means but
    physically this decomposition reflects the fact that the second law is true only {\em on average}.   For fluids made up of a non-infinite number of degrees of freedom, thermal fluctuations that decrease entropy should happen from time to time.   While, as we will say later, an operator definition of
    $\hat{T}_0^{\mu \nu}$ does exist \cite{kms,kapustagale,nishioka}, the same is not true for $\hat{\Pi}_{\mu \nu}$.
    
    Functional lagrangian hydrodynamics, based on ``doubled variables''/the Schwinger-Keldysh formalism \cite{glorioso,grozdanov,lagrangian}
    \begin{equation}
      \label{sk}
      \lnz (\phi) \rightarrow \lnz_{CTP}(\phi_+,\phi_-)\equiv \lnz (\phi_+) - \lnz (\phi_-)+\lnz_{diss}(\phi_-,\phi+) 
    \end{equation}
    with
    \[\  \frac{\delta^2 \lnz_{CTP}(\phi_-,\phi_+,)}{\delta J(t_1) \delta J(t_2>t_1)}=\frac{\delta^2 \lnz_{CTP}(\phi_+,\phi_-,)}{\delta J(t_2) \delta J(t_1)}\]
    is fundamentally inadequate to address this issue since the future time direction is defined ``at the level of the effective action'' as moving ``towards equilibrium''.  Hence, while there will be stochastic fluctuations in $T_{\mu \nu}$ any fluctuation will by definition, not affect the local increase in entropy.   This is inherent in the definition of the effective Schwinger-Keldysh action based on coarse-graining \cite{grozdanov}.

In fact, Mathematicians have known for a long time that hydrodynamics as an effective theory in terms of coarse-graining  hides ambiguities.
The existence of so-called ``wild'' or ``nightmare'' weak solutions \cite{wild} to the non-relativistic Navier-Stokes equations, the lack of uniqueness with coarse-graining (``weak solutions'') \cite{vicol,vicol2,vicol2d}, and the ``zeroth law of turbulence''/anomalous energy dissipation \cite{zeroth} shows that care must be taken with defining hydrodynamics in terms of coarse-graining and forgetting the microscopic degrees of freedom.

As physicists rather than mathematicians our interest in these formal ambiguities is limited as to what they can tell us about the physical world;  
In this spirit, it gives us the opportunity to reflect that the ``equation of state'', taken by fluid dynamicists as a parameter, is actually not a fundamental object and is directly related, via the partition function, to hydrodynamic fluctuations.
To a statistical physicist, a box of still fluid is characterised by a partition function, the maximization of entropy subject to constrains of conserved quantities.   This is what we generally call ``global equilibrium''.
Fluid dynamicists, in contrast, see the same box as a solution subject to an infinite number of possible perturbations, which then evolve within a dynamics dictated around ``local equilibrium'' (equilibrium in each fluid cell) whose stability is not strictly proven.
As also argued at the end of the appendix, these two pictures are not fully consistent.   In fact, at vanishing viscosity local equilibrium is {\em instantaneous} while global equilibrium is {\em never} achieved.   There is no limit where one picture smoothly goes into the other.
We speculate that resolving this contradiction could shed light of all the questions examined in the preceding paragraphs.

In this work we propose to do so by putting together two different approaches:  Zubarev's hydrodynamics \cite{zubarev}, which permits us to write down ideal hydrodynamics as a statistical mechanics partition function with a continuous field of Lagrange multipliers.    Crooks fluctuation theorem \cite{crooks} permits us to define an extension of Eq. \ref{piTdef} and its coarse-graining in terms of operator links, in a way reminiscent of the Wilson loop technique in quantum field theory \cite{peskin} and analogous to extensively studied quantum statistical systems coupled to heat baths  \cite{landi,landizub,zubpol,palermo,prokhorov,becstat,zhang,hamweak1,hamweak2}.

In the rest of the paper, we will formally implement a decomposition and dynamics of Eq. \ref{piTdef} via the following procedure, which can be implemented on a lattice.  The procedure can be summarised as
    \begin{description}
    \item[Take an ensemble] of configurations of the energy momentum tensor, $\hat{T}_{\mu \nu}$
    \item[Find a field $\beta_\mu$] whose Zubarev partition function approximates $\hat{T}_{\mu \nu}$.  Call the ensemble of of energy-momentum tensors generated by $\beta_\mu$ as $\hat{T}_0^{\mu \nu}$.
    \item[Construct an ensemble] of $\hat{\Pi}^{\mu \nu}=\hat{T}^{\mu \nu}- \hat{T}_0^{\mu \nu}$
    \item[Use Crooks fluctuation theorem and Gravitational Ward identities] to model the further evolution of $\hat{T}_0^{\mu \nu},\hat{\Pi}^{\mu \nu}$ as ensembles.
      \item[Read off] The resulting ensemble of Eq. \ref{piTdef} at a later time
      \end{description}
Such a procedure would allow us to evolve an initial ensemble of $\hat{T}_{\mu \nu}$ in a way that, given the assumption of approximate local equilibrium, all fluctuations are carried over.
    The next two sections will describe how this works in detail
\section{\label{eqsection}Local equilibrium: Zubarev hydrodynamics}
  In this picture, we consider a locally equilibrated fluid moving through some proper time foliation via the time-like $t$ and space $x,y,z$ coordinates
\[\  \Sigma_\mu(\tau)= (t(x,y,z,\tau),x(\tau),y(\tau),z(\tau)) \]
with the future-pointing volume element can be obtained via Stokes's theorem. 
\begin{equation}
  \label{foldef}
d\Sigma_\mu = \epsilon_{\mu \alpha \beta \gamma} \frac{\partial^\alpha t}{\partial x} \frac{\partial^\beta t}{\partial y}  \frac{\partial^\gamma t }{\partial z} dxdydz \eqcomma n_\mu \propto d\Sigma_\mu \eqcomma n_\mu n^\mu=-1 \eqcomma dz_\mu =n_\mu d\tau
\end{equation}
 Usually hydrodynamics is defined as the evolution of the average of the energy momentum tensor, which is at least approximately close to its thermodynamical equilibrium expectation value w.r.t. the frame defined by the flow vector
\begin{equation}
  \label{flowdef}
  \beta_\mu \beta^\mu = -T^{-2} \eqcomma   u_\mu=T\beta_\mu  \eqcomma u_\mu u^\mu = -1
\end{equation}
the expectation value of the energy momentum tensor is then
\begin{equation}
\label{tdef}
  \ave{T_{\mu \nu}} = \underbrace{ (e + p(e) ) u_\mu u_\nu + p g_{\mu \nu}}_{\ave{T_0}^{\mu \nu}} + \ave{\Pi_{\mu \nu}}
\end{equation}
However, thermodynamics tells us that in an equilibrium configuration $\ave{T_0}^{\mu \nu}$ is merely the most likely state and fluctuations are determined by a probability distribution given by a partition function, or equivalently an operator.
 
Let us therefore assume that the density matrix of a full quantum field, in the basis\footnote{note that in general this is not a complete basis} of the stress-energy tensor $\hat{\rho}_{T_{\mu \nu}}$ is ``close to equilibrium'' w.r.t. some flow $\beta_\mu$.   We can therefore separate the equilibrium and a non-equilibrium part at the level of the density matrix
\begin{equation}
\hat{\rho}_{T_{\mu \nu}} =\frac{ \hat{\rho}_{T_0} + \hat{\rho}_{\Pi_0}}{\mathrm{Tr} \left(\hat{\rho}_{T_0} + \hat{\rho}_{\Pi_0}  \right)} \simeq  \hat{\rho}_{T_0} \left( 1 + \delta \hat{\rho} \right)
  \label{kmscond}
\end{equation}
and the equilibrium part is given by the functional of the field $\beta_\mu$ and foliation $\Sigma_\mu$
\begin{equation}
\label{rhodef}
\hat{\rho}_{T_0}(T^{\mu \nu}_0(x),\Sigma_\mu,\beta_\mu) = \frac{\hat{e}}{Z(\Sigma_\mu,\beta_\mu)}\eqcomma Z = \mathrm{Tr}\left[ \hat{e} \right]  \eqcomma \hat{e}= \exp \left[ - \int_{\Sigma(\tau)} d \Sigma_\mu \beta_\nu \hat{T}^{\mu \nu}_0    \right]
\end{equation}
here $\hat{T}^{\mu \nu}_0$ is the {\em equilibrium} part of the energy-momentum tensor, defined at the {\em operator} level and $\beta_\mu$ is a field of Lagrange multiplies.    Physically, the definition of equilibrium via Eq. \ref{kmscond} and Eq. \ref{rhodef} means {\em all} moments rather than just the average can be calculated from Eq. \ref{rhodef} and derivatives of the partition function w.r.t. $\beta_\mu$.
``separating'' a density matrix and defining equilibrium at the density matrix level looks unfamiliar, but it is strictly speaking possible from the partition function
\begin{equation}
  \label{kmscondz}
  Z = Z_{T_0} \times Z_\Pi
  \end{equation}
, as is explicitly shown in appendix \ref{appendixz}.  It is similar to the definition of $\hat{\rho}_{les}$ in \cite{landizub} for a quantum system with a finite number of degrees of freedom.    Of course in quantum field theories density matrices are not consistently normalized, but as we will see this problem might be irrelevant in the dynamics.

Eqs. \ref{kmscond},\ref{rhodef} and \ref{piTdef} can be related to Eq. \ref{hyerarchyscales} via the {\em ergodic hypothesis} \cite{bass}, which says that a probability distribution sampled over time $\mathcal{P}(t)$ can be approximated via a state average $\mathcal{P}(\mu)$
\begin{equation}
  \label{ergodic}
    \ave{\hat{O}\mathcal{P}(t)}_t = \ave{\hat{O}\mathcal{P}(\mu)}_\mu
  \end{equation}
The second, thermalization inequality in Equation \ref{hyerarchyscales} is satisfied when the timescale when Eq. \ref{ergodic} is a good approximation is parametrically smaller than the macroscopic evolution scale.   The first inequality is satisfied when the ensembles are equivalent, i.e., for a generic observable $\hat{O}$
\begin{equation}
  \label{thermolym}
  \mathcal{P}(\mu) \rightarrow \delta(\mu - \ave{\mu})
  \end{equation}
The ``small fluid limit'', therefore, is when Eq. \ref{ergodic} holds but Eq. \ref{thermolym} does not.   Hence, we must consider equilibrium to be a {\em statistical operator} rather than {\em its average}.   The generating function equation \ref{kmscondz}, expanded in the appendix, does exactly this.

Note that only in full local equilibrium {\em and} irrotational flow ($n_\mu \propto \beta_\mu$) can $\hat{\rho}_{\Pi}=0$.
Otherwise the choice of $\beta_\mu$ and $\Sigma_\mu$ is of course somewhat arbitrary, just like it is in Israel-Stewart hydrodynamics (where it leads to the definition of $\Pi_{\mu \nu}$) and the Hamiltonian of weak force effective theory \cite{hamweak1,hamweak2}.   We need it ``close enough to equilibrium'' ($\rho_{T_0}$ ``close enough'' to the full matrix) that some near-equilibrium effective theory (in our case Crooks fluctuation theorem) will be a good effective theory to calculate $\hat{\rho}_{\Pi}$.  Note that, as we can see if we use $J(x)$ to construct a smeared test function, the well-posedness of a coarse-grained Eq. \ref{denspart} is intimately connected to the existence of weak solutions, which mathematicians are still discussing \cite{vicol,vicol2}.   This underscores the importance of defining hydrodynamic quantities at operator level.
 
However, this definition of $\hat{\rho}$ in terms of $\hat{T}_0^{\mu \nu}$ is incomplete, since deviations from equilibrium are left out.   It also is ``covariant'' but time evolution is not included.   Previous approaches (see \cite{zubarev} and references therein) treat non-equilibrium processes as coarse-graining of
$\ave{T^{\mu \nu}-T_0^{\mu \nu}}$ and derive dynamics from a gradient expansion of $\ave{T}_{\mu \nu}$ and conservation laws.   As mentioned in the introduction, this approach generally breaks causality and there is no clear fluctuation-dissipation relation.   In the next section we will argue that Crooks fluctuation theorem provides an alternative formulation that obviates this difficulty.
\section{Deviations from equilibrium via Crooks fluctuation theorem}
\subsection{Non-equilibrium non-fluctuating hydrodynamics}
Our purpose is to find a non-equilibrium fluctuation dynamics for $\hat{\Pi}_{\mu \nu}$.   The density matrix of Eq. \ref{rhodef} should not depend on it for the KMS condition to hold, and  it should be ``subleading'' and determined entirely from local variations around the equilibrium part.

In  Standard treatments of hydrodynamics characterized just by expectation values, therefore,  $\Pi_{\mu \nu}$ (either via a gradient expansion, as in Navier-Stokes, or as independent degrees of freedom which relax to its equilibrium value, as in Israel-Stewart \cite{kodama}) is determined via the second law of thermodynamics, the non-decrease of entropy with proper time .  Entropy can be formulated microscopically via the Von Neumann entropy definition \cite{zubarev} and  Eq. \ref{rhodef}
\begin{equation}
\label{entropylim}
s = -Tr (\hat{\rho} \ln \hat{\rho}) = -\frac{d}{dT}
\left( T\lnz\right) 
\end{equation}
Given that entropy is maximized at equilibrium, the second law can lead to an average definition of entropy close to equilibrium \cite{romentropy}
\begin{equation}
  \label{secondlaw}
  n^\nu \partial_\nu \left( s u^\mu \right) = n^\mu \frac{\Pi^{\alpha \beta}}{T} \partial_\alpha \beta_\beta \geq 0
  \end{equation}
Equation \ref{entropylim} and \ref{rhodef} means that the entropy is determined entirely from equilibrium as well as the foliation vector $n_\mu$.  Conversely, a ``good choice'' of $n^\mu$ should respect Eq. \ref{secondlaw} according to the thermodynamic arrow of time \cite{zubarev}.  If hydrodynamics is a good effective theory, then, $\Pi_{\mu \nu}$ is deducible from the long-time behaviour of the correlator \cite{kadanoff}
\begin{equation}
  \label{kubo}
\partial_\mu (s u^\nu) \simeq \left[ \partial^\alpha u^\beta \right] \times \lim_{w \rightarrow 0} \frac{1}{w}\mathrm{Im} \ave{ \left[ \tilde{T}_{\alpha \mu}(w) \tilde{T}_{\beta \nu}(0)  \right]}
  \end{equation}
However, some issues remain.  In the Landau frame, one usually has to assume the transversality condition $u_\mu \Pi^{\mu \nu}=0$ to preserve 
$u_\mu$ as the Killing vectors of the foliation metric (in physical language, to distinguish non-equilibrium from advective Heat flow).
However, this can be done only  provided $\Sigma_\mu \propto u_\mu$, and in situations where vorticity is relevant this is impossible globally.  More importantly, equation \ref{secondlaw} does not take thermal fluctuations and higher cumulants into account.   

Given the limits summarized here, we propose to fix $\hat{\Pi}_{\mu \nu}$ in Eq. \ref{piTdef} by using Crooks theorem Eq. \ref{crooksth} as a ``dynamical update'' for fluctuating hydrodynamics.
\subsection{Fluctuation and dissipation via Crooks theorem in Quantum Mechanics}
Crooks's theorem \cite{crooks} is a principle that relates the ''probability of a work configuration being done in reverse'' (denoted by $-W$) to the probability of work being done ``as usual'' (denoted by $W$) to the entropy produced by it $\Delta S(W)$
\begin{equation}
  \label{crooksth}
   \frac{P(-W)}{P(W)} = \exp[-\Delta S(W)]
\end{equation}
It is a powerful tool since it is valid far from equilibrium, being dependent for its validity on the existence of an equilibrium state {\em somewhere} in the phase space, microscopic time reversibility and Markovian evolution.   In this section, let us summarize how this works within quantum mechanics, taking \cite{landizub} as an example.

In this work, the quantum mechanics of a system in this limit has been derived as the near equilibrium stationary state (NESS) perturbed from a local equilibrium state (LES) by a ``kick'' in the parameter space $\lambda$ (we omit chemical potential terms for simplicity)
\begin{equation}
  \label{zuberdef}
\hat{\rho}_{ness} = \hat{\rho} (\lambda+\delta \lambda) \simeq \hat{\rho}_{les}(\lambda) e^{\hat{\Sigma}} \frac{Z_{les}}{Z_{ness}} \eqcomma \hat{\rho}_{les}=\frac{1}{Z_{les}} \exp\left[-\frac{\hat{H}}{T}  \right]
\end{equation}
where ``les'' is equivalent to the local equilibrium  state (the similarity with Eq. \ref{rhodef} is obvious) and ``ness'' the near-equilibrium stationary state (this is a quantum system, so $\hat{\rho}$ only depends on a finite set of degrees of freedom and time).
where $\hat{\Sigma}$ is an operator whose expectation gives the entropy production rate.  The correctly normalized $\hat{\rho}_{les,ness}$ can be obtained from $Z_{les,ness}$ via Eq. \ref{denspart} in 0+1 dimensions.     Crook's theorem permits in principle to close Eq \ref{zuberdef} in operator form since $\hat{\Sigma}$ is connected to the Hamiltonian via a Kubo-like relation \cite{kadanoff}.   In the absence of chemical potentials, this would be
\begin{equation}
  \label{sigmakubo}
\hat{\Sigma} = \delta_{1/T} \Delta \hat{H}_+ \eqcomma \hat{H}_+ = \lim_{\epsilon\rightarrow 0^+} \epsilon \int dt e^{i \epsilon t} e^{-i\hat{H}t} \Delta \hat{H} e^{i\hat{H}t} 
\end{equation}
where $\hat{H}$ is the full evolving Hamiltonian and $\Delta \hat{H}$ represents the difference of the hamiltonians between two reservoirs.
Equivalently, any correlation and entanglement between the equilibrium and non-equilibrium part of $\hat{\rho}$ is taken care of by the evolution of $\hat{\Sigma}$.    

This approach was then used to derive uncertainity relations \cite{landi,landizub,zhang} of the form
\begin{equation}
  \label{heatcap}
\frac{   \ave{(\Delta Q)^2}}{\ave{Q}^2} \geq \frac{2}{\Delta S(W)} \Rightarrow 
\frac{d}{d\tau} \Delta S \geq \frac{1}{2} \frac{d}{d\tau} \frac{\ave{Q}^2}{\ave{(\Delta Q)^2}}
\end{equation}
However it cannot readily be translated into the Gibbsian microstate picture as the relation of a generic definition of ``work'' to microstates is lacking.
For quantum fields near to local equilibrium, however, this definition is readily given by $d \hat{W}_\nu=\hat{T}_{\mu \nu} dx^\mu$.

Let us therefore try to generalize Crooks fluctuation theorem from $0+1d$ quantum mechanics to higher dimensional field theory.
\subsection{Proposed generalization to field theory}
At first sight, the construction in the previous section looks like an arbitrary extra assumption, since it is tempting to interpret the Zubarev $\lnz$ as an effective lagrangian of a ``field of $\beta_\mu$, which means dynamics must be determined by a functional integral.
To clarify this, we recall that the KMS condition can reduce the functional integral to a form computable by a Metropolis type weighting \cite{peskin}.   We also recall that Crooks theorem is proven for Markovian systems.   In qualitative language, assuming Crooks theorem means an outcome of the ``correlation between adjacent cells in our foliation'' is determined by ``how many ways are there'' for this outcome to occur.   If the system is close to local equilibrium, this should be a good approximation.

Thus, the dynamics of Eq. \ref{crooksth} reduces to the kind of ``effective action'' one computes on the lattice, assuming each element at rest with $\beta_\mu$ is close to local equilibrium.      In this respect, the evolution of such a fluctuating fluid can be compared to the evolution, in computer time, of the system studied in \cite{jarz} (where Jarzynski's equality, equivalent to Crook's fluctuation theorem, was employed).  The relationship between the coarse-graining using Crook's theorem and the more traditional hydrodynamic gradient expansion can be seen as analogous as the relationship between the coarse-graining of QCD via Wilson loops \cite{peskin} and effective theories based on hadrons (chiral perturbation theory and so on).   The second are intuitive but effectively ``classical'' (ambiguous beyond tree level), the first has the potential to describe higher order fluctuations but is much less liable to intuition.

Our task is to try to find an analogue definition of $\hat{\Pi}$ and its density matrix given the density matrix defined in Eq. \ref{rhodef} and \ref{kmscond}.
Other than relativistic covariance, the difference is the fact that we have a field (a fluid) with cells interacting with each other rather than a quantum system with a finite number of degrees of freedom interacting with a fluctuating bath.   The fact that this is a field allows us to think in terms of hierarchy of scales, and to treat an infinitesimal work and dissipation done by neighbouring fluid cells analogously to the work and dissipation done by the two systems in \cite{landizub}.  In this respect, we note that Eq. \ref{zuberdef} critically depends on commutation between $\hat{\Sigma}$ and $\hat{H}$, which in that work is a requirement for Markovian system-bath interaction, one of the assumptions of the Crooks fluctuation theorem.       The zero commutation also arises assuming a fast decoherence between system and bath \cite{hamweak1,hamweak2}.

In this work, we are using the Crooks fluctuation ``theorem'' as a postulate, but this aspect might be problematic for its applicability, since in QFT every point is correlated with every other.     In the appendix \ref{appendixz} we will argue that the applicability of the Crooks theorem can nevertheless be justified in terms of scale separation, Eq. \ref{hyerarchyscales}: In the usual construction of hydrodynamics as an effective theory, the coefficient of this tail is related to the sound and viscosity poles, and the dynamics at the scale of the hydrodynamic gradients is assumed  to be dominated by the tail, as in Eq. \ref{kubo}. Eq. \ref{sigmakubo} parallels the Kubo formula, in that it isolates the long-time tail of a commutator.     Here, we make exactly the same assumption, but at the level of operators, which allows us to ``carry all terms'' of the first inequality in Eq. \ref{hyerarchyscales}, which, as argued in \cite{ryblewski}, acts like a ``Planck constant''
\color{black}

\begin{figure}
     \epsfig{width=0.99\textwidth,figure=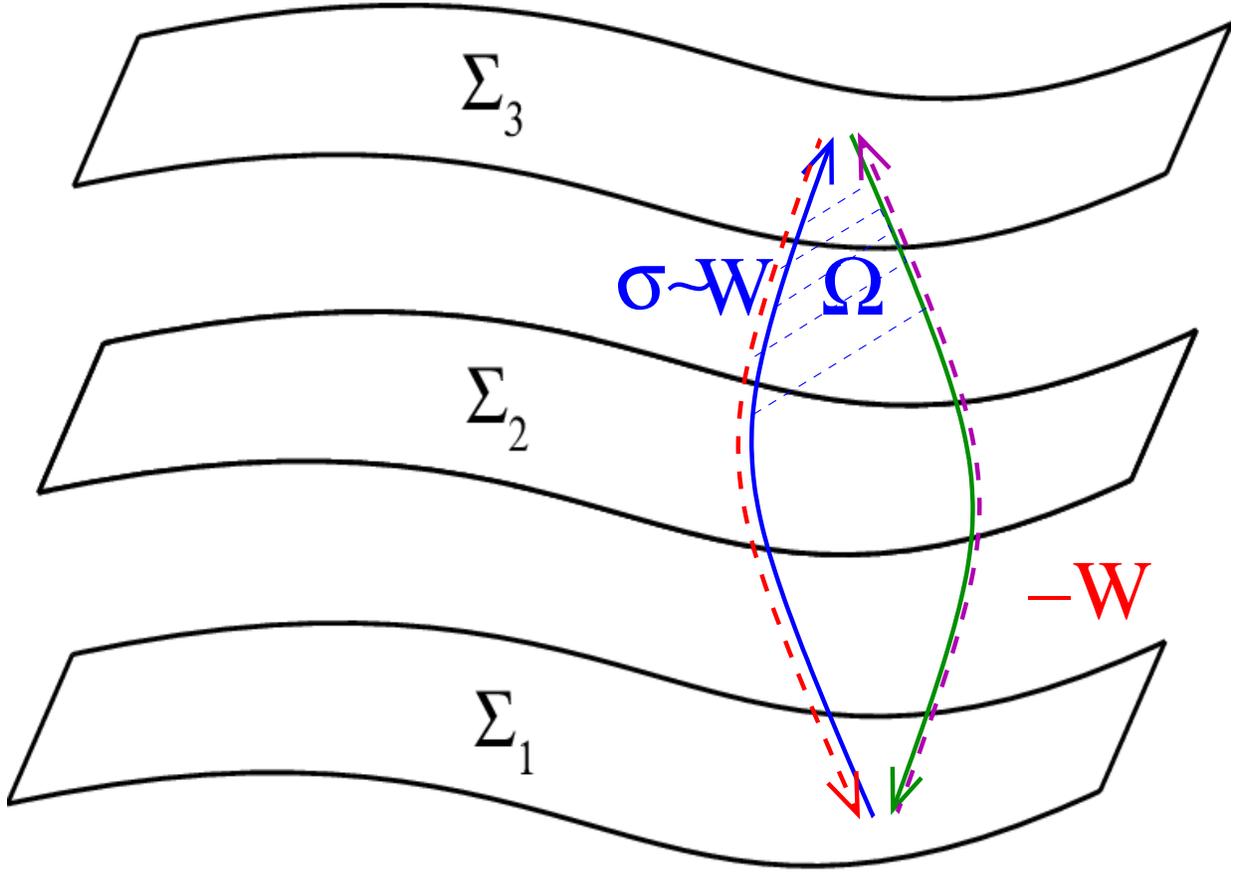}
\caption{\label{foliation}  A generic foliation $\Sigma$, with the forward (dissipation-driven) vs backward (fluctuation-driven) direction indicated by solid and dashed lines, and an infinitesimal transformation denoted by $\Gamma$.}
\end{figure}

Let us therefore consider a fluid cell travelling through a given path $d \Sigma_\mu(\tau)$ (Fig. \ref{foliation}).   One can imagine the usual, partially dissipative fluid evolution from beginning to end, where we keep track of $e,u_\mu,\Pi_{\mu \nu}$ across each segment of $d\tau$ of the path.   By Stokes's theorem we know that
\begin{equation}
\label{stokes}
  - \int_{\Sigma(\tau_0)} \!\!\!\!\!\! \di \Sigma_\mu \; \left( \wT^{\mu\nu}
  \beta_\nu 
  \right)
  =
 - \int_{\Sigma(\tau')} \!\!\!\!\!\! \di \Sigma_\mu \; \left( \wT^{\mu\nu}
 \beta_\nu 
 \right)
 + \int_\Omega \di \Omega \; \left( \wT^{\mu\nu} \nabla_\mu \beta_\nu
 \right),
\end{equation}
where $\Omega$ is a hypersurface element enclosing the two paths, in 4d a dimension of a volume (Fig. \ref{foliation}) and $\sigma(\tau)$ is a 1d path parametrizing the direction of $n^\mu$ in spacetime.

This relation is exact, since it follows from geometry.   It will hold on any field configuration in the ensemble.
One can also imagine, since we are dealing with a fluctuating thermal/quantum system, that  a random fluctuation could, with a certain probability give us {\em exactly the reverse time-evolution of this path}.   Crooks theorem, if it applies to a quantum field system close to local equilibrium, gives a constraint, relating the work done by the fluid cell in each of these situations to the entropy produced (Fig \ref{foliation} solid and dashed lines).

Regarding Eq. \ref{rhodef} as being proportional to probabilities of given paths, one can construct a ratio of probabilities by simply reversing the time direction in the first term of the RHS of Eq. \ref{stokes}, putting the second term of Eq. \ref{stokes} equal to the dissipative term, and using Eq. \ref{rhodef} to construct the probabilities. The partition function $Z$ cancels out so only the unnormalized probabilities remain, building up a conditional version of $\mathcal{P}(...)$ of Eq. \ref{factp} as
\begin{equation}
\label{hydrocrook}
\frac{\mathcal{P}\left( T_{\mu \nu}'\left(n^\mu + d\Sigma^\mu \right)|T_{\mu \nu}(n^\mu)\right)}{\mathcal{P}\left( T_{\mu \nu}\left(n^\mu + d\Sigma^\mu  \right)|T_{\mu \nu}'(n^\mu) \right)  } = \frac{\exp \left[ - \int_{\sigma(\tau)} d z_\mu \beta_\nu \hat{T}^{\mu \nu}    \right]}{\exp \left[ - \int_{-\sigma(\tau)} d z_\mu \beta_\nu \hat{T}^{\mu \nu}    \right]} = \exp \left[ \frac{1}{2} \int_{\Omega} d \Omega_{\mu}^\mu \left[ \frac{\hat{\Pi}^{\alpha \beta}}{T} \right] \partial_\beta \beta_\alpha  \right]
\end{equation}
where the different line elements are done according to Eq. \ref{foldef}. $\sigma(\tau)$ is a path of a volume element moving according to some foliation and $d\Omega$ is the future-oriented surface integral between the two loops.   Note that this is a {\em ratio} of probabilities, hence divergences of $Z$ which affect the density matrix Eq. \ref{nishioka} cancel out.

This equation should be true for any foliation, and relate the global evolution of the volume element (the left hand side) to the entropy change through this evolution (the right hand side).   Note that if $\hat{T}_{\mu \nu}$ is constructed out of a sampling of particles, its non-relativistic linear limit trivially reproduces the results in the fluctuating hydrodynamics of \cite{csernai}, via a Taylor-expansion and the averages of \ref{hydrocrook}. However, Eq. \ref{hydrocrook} is defined non-perturbatively.   

Equation \ref{hydrocrook} relates, given an arbitrary foliation $\Sigma_\mu$, the fluctuations in $u_\mu$ to the fluctuations in $\Pi_{\mu \nu}$ {\em defined in that foliation}.
Since this equation is separately valid for any path $\sigma(\tau)$, moving forward or backward, we should be able to deform the path by $\Gamma$ (cyan arrow in Fig. \ref{foliation} and get a similar couple of paths.  Basic differential geometry allows any two foliations $d\Sigma_\mu$ and $d\Sigma_\nu$ to be related by
\begin{equation}
  \label{transport}
  d \Sigma'_\mu = d\Sigma_\mu + Q_\mu^\nu d\Sigma_\nu \eqcomma Q_{\mu \nu} = \partial_\alpha \Gamma^\alpha_{\mu \nu} + \Gamma^{\alpha \beta}_{\mu } \Gamma_{\alpha \beta \nu} \eqcomma \omega_{\mu \nu} = g_{\mu \nu} - Q_{\mu \nu}
\end{equation}
with the latter definition defining a transport along a certain foliation forward and a slightly different path backward.   We can also, in the RHS, parametrize $d \Omega_{\mu}^\mu = d\Sigma_\mu \beta^\mu$ according to co-ordinates co-moving with $\beta_\mu$.
Equation \ref{hydrocrook} then becomes 
\begin{equation}
\label{hydroloop}
\ave{ \exp \left[  \oint d z_\mu \omega^{\mu \nu} \beta^\alpha \hat{T}_{\alpha \nu}    \right]} = \ave{ \exp \left[ \int \frac{1}{2} d \Sigma_\mu \beta^{\mu} \hat{\Pi}^{\alpha \beta} \partial_\alpha \beta_\beta  \right]}
\end{equation}
Note that $\omega_{\mu \nu}$ is a closed path (looping forward for one choice os $\sigma$ and backward for another), while $d\Sigma_\mu$ is a volume.

If we coarse-grain to short intervals and loops, we can Taylor-expand Eq. \ref{hydroloop} and also take the Gaussian approximation (only two-point correlators matter), which is usually an assumption required for the Crooks fluctuation theorem \cite{crooks}.   In this case, Eq. \ref{hydroloop} is directly connected to the propagator of $T_{\alpha \beta}$
\begin{equation}
  \label{gaussian}
 \omega^{\mu \nu} \beta^\alpha \ave{ \left[ \hat{T}_{\mu \nu} \hat{T}_{\alpha \beta} \right]}= \frac{1}{2}\beta^\gamma \frac{\partial \Sigma_{\gamma}}{\partial x^\beta}\ave{ \hat{\Pi}_{\mu \nu}} \partial^\mu \beta^\nu
\end{equation}
The commutator in the above equation, of course, contains the microscopic Kubo formulae used to derive shear and bulk viscosity \cite{kodama}.  However, it also contains microscopic fluctuations of statistical mechanics, and treats them along the same footing.   Equation \ref{rhodef} permits both to be encoded in the partition function.
Indeed, an inversion of Eq. \ref{secondlaw} together with Eq. \ref{piTdef}  can be used as a {\em definition} of $\hat{\Pi}_{\mu \nu}$ as an operator connecting two fluid cells across an element of foliation
\begin{equation}
  \label{crookspidef}
 \left. \frac{\hat{\Pi}^{\mu \nu}}{T}\right|_{\sigma} = \left( \frac{1}{\partial_\mu \beta_\nu} \right) \frac{\delta}{\delta \sigma} \left[  \int_{\sigma(\tau)} d \Sigma_\mu \beta_\nu \hat{T}^{\mu \nu} - \int_{-\sigma(\tau)} d \Sigma_\mu \beta_\nu \hat{T}^{\mu \nu}    \right]
\end{equation}
\subsection{Particular cases and limits}
We can do some sanity checks, as shown in Fig. \ref{contours}.
A purely timelike, $\sigma \propto t, d\Sigma_\mu = (0,dt \vec{\nabla} \times \vec{x})$ the application of Eq. \ref{crookspidef} on a hydrostatic background straight-forwardly reproduces $\Pi_{\mu \nu} \propto \eta$ given by the Kubo formula Eq. \ref{kubo} (Eq. \ref{crookspidef} is basically the Kubo formula in operator form).
The opposite spacelike limit $d\Sigma_\mu=(dV,\vec{0})$ (Fig. \ref{contours} left panel) the above definition and Eq. \ref{hydrocrook} recover the Boltzmann entropy relation from statistical mechanics 
\begin{equation}
  \label{statentropy}
  \frac{\Pi_{\mu \nu}}{T} u^\mu d\Sigma^\nu \rightarrow\Delta S= \frac{\Delta Q}{T} =
  \ln\left( \frac{N_1}{N_2} \right)
\end{equation}
where $N_{1,2}$ are the number of microstates (coming in a ratio, which regularizes the divergence).

When viscosity goes to zero {\em and} the particle density goes to infinity 
(the first two terms in the hierarchy Eq. \ref{hyerarchyscales}), 
Crooks fluctuation theorem gives $P(W) \rightarrow 1$ $P(-W) \rightarrow 0$ $\Delta S \rightarrow \infty$ 
so Eq. \ref{hydrocrook} reduces to $\delta$-functions of the entropy current
\begin{equation}
  \label{deltasid}
 \delta \left(d\Sigma_\mu \left( s u^\mu \right) \right) =0
\end{equation}
We therefore recover conservation equations for the entropy current, which without chemical potentials define ideal hydrodynamics.

Our approach therefore reproduces hydrostatics and ideal hydrodynamics.
For those situations where neither $\alpha$ nor $K$ in Eq. \ref{hyerarchyscales} are negligible, so the $\delta-$function in Eq. \ref{deltasid} becomes smeared out non-linearly in a way related to $\hat{\Pi}_{\mu \nu}$.   Physically, this can either occur in the deep turbulent regime, or in the vicinity of the critical point where microscopic fluctuations diverge \cite{stephcrit}.  
\subsection{The equations of motion for the partition function}
Comparing these formulae to those below Eq. \ref{zuberdef}, it is clear that Eqs \ref{rhodef} and \ref{crookspidef} fulfill the role we wanted, in that they are close analogues of the ``local equilibrium state'' $\hat{\rho}_{les}$, the transition between them via an entropy operator can be reduced to a Kubo-like formula \cite{kadanoff} and they maintain the Lorentz symmetry and the isotropic symmetry and KMS condition of local equilibrium at the operator level, analogously to Wilson lines in QCD.    This could open the way to a functional differential equation in terms of the partition function rather than just equations of motion for the averages.

To obtain a solvable set of equations, we need to understand what $\Gamma_{\alpha \beta \gamma}$ look like for coordinates defined by $n_\alpha$.   Remembering Eq. \ref{foldef} the microscopic transformations are
\begin{equation}
  \label{nalpha}
g_{\mu \nu} \rightarrow g_{\mu \nu} + \partial_\mu z_\nu + \partial_\nu z_\mu \eqcomma  d\tau = n_\mu dx^\mu \eqcomma d\Sigma_\mu = n_\mu dV  d\tau = dV dz^\mu
\end{equation}
which immediately means that
\begin{equation}
  \Gamma_{\alpha \beta \gamma} = \frac{1}{2} \left( A_{\alpha \beta \gamma} + A_{\alpha \gamma \beta} - A_{\beta \alpha \gamma} \right) \eqcomma A_{\alpha \beta \gamma} = \partial_\alpha \partial_\beta z_\gamma
\end{equation}
\begin{equation}
  \label{omegadef}
  Q_{ \beta \gamma} = \frac{1}{2} \left( \partial^2 \left(\partial_\gamma z_\beta+\partial_\beta z_\gamma \right) - \partial_\beta \partial_\gamma \left(\partial.z
  \right) + \partial^2 z_\beta \partial^2 z_\gamma + A_{\mu \nu \beta} A^{\mu \nu}_\gamma  - A_{\beta \mu \nu} A^{\mu \nu}_{\gamma} + A_{\beta \mu \nu} A_{\gamma}^{ \mu \nu}  \right) 
\end{equation}
\begin{figure*}
  \begin{center}
    \epsfig{width=0.99\textwidth,figure=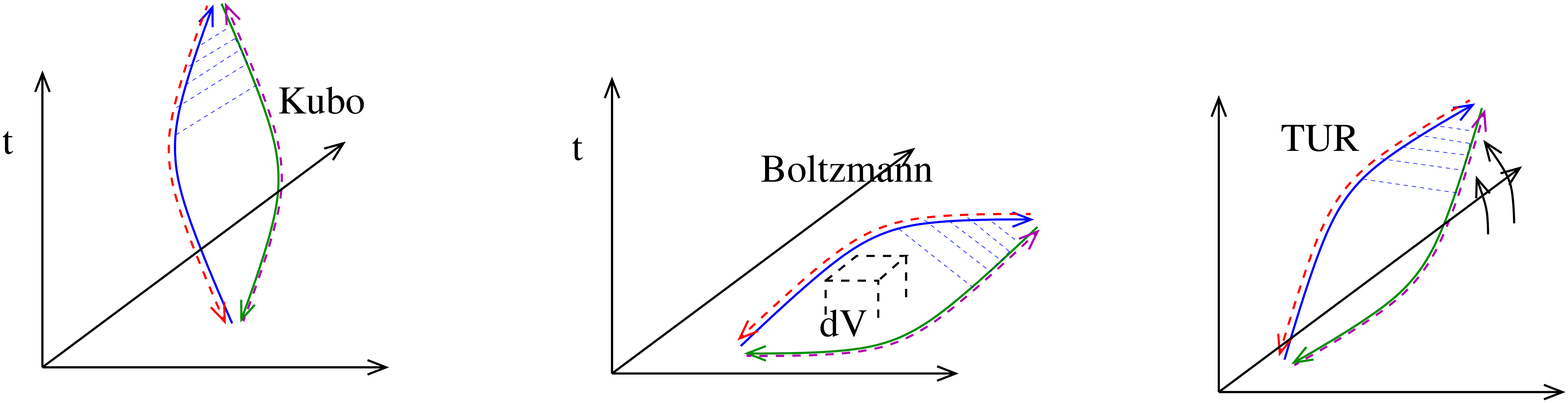}
    \end{center}
  \caption{\label{contours} Left panel:  A purely time-oriented contour, reproducing the Kubo formula
    Center panel: A space-like contour, expected to reproduce Boltzmann entropy contour.  Right panel:  The same contour slightly tilted in time, expected to produce hydrodynamic uncertainity relations and anomalous energy dissipation Eq. \ref{tur}.
       }
\end{figure*}
Because of Lorentz invariance, any such an infinitesimal transformation is generated by $T_{\mu \nu}$.  This gives rise to Ward identities, \cite{boulware,jeon,kadanoff} linking the propagator of $T_{\mu \nu}$ to its expectation values.
\begin{equation}
  \label{ward}
\partial^\alpha \left\{ \ave{\left[ \hat{T}_{\mu \nu}(x),\hat{T}_{\alpha \beta} (x') \right]} - \delta(x-x') \left( g_{\beta \mu} \ave{\hat{T}_{\alpha \nu}(x')} + g_{\beta \nu} \ave{\hat{T}_{\alpha \mu}(x')}  - g_{\beta \alpha} \ave{\hat{T}_{\mu \nu}(x')}  \right) \right\} =0 
\end{equation}

To proceed further, we use the Zubarev partition function defined in Eq. \ref{rhodef}. We can do this using the entropy definition and the assumption that the {\em equilibrium} part of the energy momentum tensor is exact and not an average (note that unlike in \cite{becstat} the stress-energy tensor is not totally in equilibrium, its just that the equilibrium part includes fluctuations).    This allows us to take higher order derivatives.   For example, the energy-momentum tensor expectation value, which by Lorentz invariance coincides with the functional derivative of $\lnz$ w.r.t. the metric \cite{peskin,sonhydro}, can be rewritten as the sum of the equilibrium part $T^{\mu \nu}_0$ and the non-equilibrium part $\Pi^{\mu \nu}$
\begin{equation}
\label{zpartdiss}
\ave{T_{\mu \nu}} = \frac{2}{\sqrt{-g}}\frac{\delta \lnz}{\delta g^{\mu \nu}} = \ave{T_0}^{\mu \nu} + \ave{\Pi}^{\mu \nu}
\end{equation}
where
\begin{equation}
  \label{t0def}
  \ave{T_0^{\mu \nu}} =\frac{ \delta^2 \lnz}{\delta \beta_{\mu} dn_\nu}
\end{equation}
\begin{equation}
  \label{pidef}
\ave{  \Pi^{\mu \nu}}= \frac{2}{\partial_\mu \beta_\nu} \partial_\gamma \frac{\delta}{\delta \ln (\beta_\alpha \beta^\alpha)} \left[  \beta^{\gamma} \lnz \right]
\end{equation}
here, the first term comes from the definition of the equilibrium density matrix and the second is a straight-forward algebraic manipulation of Eq. \ref{entropylim}.   Note that Eq. \ref{pidef} is not necessarily perpendicular to flow, since 

Eq. \ref{ward} can be rewritten as
\begin{equation}
  \label{wardpart}
  \partial_\alpha  \left[ \frac{2}{\sqrt{-g}} \frac{\delta^2 \lnz}{\delta g_{\mu \nu} \delta g_{\alpha \beta}} - \delta(x-x') \frac{2}{\sqrt{-g}} \left( g_{\beta \mu} \frac{\delta \lnz}{\delta g_{\alpha \nu} }
  + g_{\beta \nu} \frac{\delta \lnz}{\delta g_{\alpha \mu} }  - g_{\beta \alpha} \frac{\delta \lnz}{\delta g_{\nu \mu} }    \right) \right]  =0 
\end{equation}
and, finally, Crook's theorem Eq. \ref{hydroloop}, combined with Eq. \ref{omegadef},Eq. \ref{t0def}, Eq. \ref{wardpart} becomes
\begin{equation}
  \label{crookszeta}
  \exp \left[ 2 \oint d x_\mu \omega^{\mu \nu} \beta^\alpha  \frac{\delta \lnz}{\delta g^{\mu \nu}}   \right] = \exp \left[ \int_{\sigma(\tau)} d \Sigma_\mu \beta^{\mu}  \partial_\gamma \frac{\delta}{\delta \ln (\beta_\alpha \beta^\alpha)} \left[  \beta^{\gamma} \lnz \right]
       \right]
\end{equation}
This equation can be used as a basis of the Metropolis algorithm described in the next section.   However it can also be expanded using Eq. \ref{gaussian} as
\begin{equation}
  \label{zgaussian}
  \frac{\delta^2}{\delta g^{\mu \nu} \delta g^{\alpha \beta}} \lnz
 = \sqrt{-g} \frac{ \beta_\kappa}{ \omega^{\mu \nu} \beta^\alpha} \partial_\beta n^\kappa  \partial_\gamma \frac{\delta}{\delta \ln (\beta_\alpha \beta^\alpha)} \left[  \beta^{\gamma} \lnz \right]
\end{equation}
These are three equations, with three unknowns for each point in space, the components $\beta_\kappa$. The Ward identity Eq. \ref{wardpart} brings the number of independent components to three by ensuring that there exists a local Lorentz transformation $\Lambda_{\mu \nu}$ such that
\[\ \beta_\nu = \Lambda_{\mu \nu} \beta_0^\mu \eqcomma \beta_0^\mu = (1/T,\vec{0})  \]
and the $\beta_\mu$ that regulates Eq. \ref{rhodef} is the same that weights $\hat{\Pi}^{\mu \nu}/T$ in such a way as to ensure that the energy conservation equation Eq. \ref{piTdef} is satisfied.

As is well-known, any velocity field $u_\mu$ can be decomposed into an unvortical potential part and a vorticity part $\zeta^\mu$, and the unvortical part can be written as a potential $\phi$.  Stokes's theorem precludes $n_\mu$ to be proportional to vorticity.   This means a good choice is
\begin{equation}
  \label{alphabeta}
\beta_\mu = \partial_\mu \phi + \zeta_\mu \eqcomma n_\mu \rightarrow T \partial_\mu \phi \eqcomma \omega_{\mu \nu} = g_{\mu \nu}
\end{equation}
These equations together  with the Ward identity Eq. \ref{ward} 
define the equations of motion of $T_{\mu \nu}$ and its propagator ``non-perturbatively'' close to local equilibrium.   This can be argued to be the foliation that most respects equilibrium.   In general, however, any vorticity present will ensure this foliation is never strictly at equilibrium and the RHS of Eq. \ref{crookszeta} and \ref{zgaussian} do not vanish, producing dissipation.  In light of Eq. \ref{alphabeta}, choosing $n_\mu \propto \partial_\mu \phi$ will make gradients of velocity cancel out in Eq. \ref{gaussian}.   In analogy with \cite{landi}
a thermodynamic uncertainity relation will relate reversibility (as parametrized by the commutator) and thermodynamic fluctuations to the inverse of the entropy projected in the direction of vorticity.  The resulting thermodynamic uncertainity relation is therefore conjectured to have this form, similar to Eq. \ref{heatcap}
\begin{equation}
  \label{tur}
\frac{\ave{\left[ T_{\mu \gamma}, T_{\nu}^{ \gamma} \right]}}{\ave{ T^{\mu \nu}}^2} \geq \frac{ \mathcal{C} \epsilon_{\mu \gamma \kappa} \ave{T^{\gamma \kappa}} \beta^\mu  }{ \Pi^{\alpha \beta} \partial_\beta \zeta_\alpha} \eqcomma \mathcal{C}\sim \order{1}
\end{equation}
One can understand the qualitative form of this equation by making a link to turbulence \cite{giorgio}:   As the microscopic viscosity decreases, the system becomes more turbulent.  This means that thermal fluctuations will increasingly be converted in hydrodynamic modes, with a random source $\sim \ave{\left[ T_{\mu \gamma} T_{\nu}^{ \gamma} \right]}$
In 3D, this is accompanied by growth of microscopic vorticity $\sim \epsilon_{\mu \gamma \kappa} \ave{T^{\gamma \kappa}} \sim \zeta_\mu$ (Eq. \ref{alphabeta}), which cannot be foliated reversibly.  Hence, the RHS of Eq. \ref{crookspidef} will diverge because of the $\partial_\mu \beta_\nu$ term in the denominator.
The result is a minimum-setting relationship between viscosity, vorticity and thermal fluctuations of the form given in Eq. \ref{tur} which has the potential to explain the ``Zeroth law'' \cite{zeroth}.    Equilibrium thermodynamic fluctuations in tandem with microscopic deviations from potential flow set the minimum of anomalous dissipation.   Note that in 1 and 2D, where $\partial_\mu \zeta_\nu$ does not exist or is restricted, we know that anomalous dissipation is absent.

We can obtain a further insight into $\mathcal{C}$ by looking at Fig. \ref{contours} right panel in comparison to the other two panels.   Tilting the contour a little bit in the time direction is equivalent to going from purely space-like fluctuations (described by the second-order derivative of the finite temperature partition function) to an infinitely weak fluctuation that decays in time.   In other words, one expects that Eq \ref{tur} goes between the limit of thermal fluctuations in the flat contour case (left-hand panel of Fig. \ref{contours}) to the Kubo formula (right-hand panel of Fig. \ref{contours}).   This is possible if
\begin{equation}
  \label{cw}
 \mathcal{C} = \lim_{w \rightarrow 0} \frac{\mathrm{Re}\left[F(w)\right]}{\mathrm{Im}\left[F(w)\right]}
 \eqcomma F(w) = \int d^3 x dt \ave{T^{x y}(x) T^{x y}(0)} e^{i(kx-wt)}
\end{equation}
This allows an experimental test of the picture presented here, since Eq. \ref{cw} gives a quantitative prediction for how the ``zeroth law of turbulence'' develops.    Note that most of the mathematical literature on the topic \cite{zeroth} assumes {\em incompressible} fluids, for which there is no relativistic continuation.   One would have to develop a non-relativistic limit of Eq. \ref{tur}, something beyond in scope of this current work.
\subsection{The dynamics}
We are now ready to try to make sense what we derived.
The evolution of the energy momentum tensor will be given, in the Gaussian approximation, via a foliation $d\Sigma_{\mu}$ and a commutator,  as  a stochastic $it\hat{o}$ integral \cite{bass} 
\begin{equation}
  \label{itoeq}
  \hat{T}_{\mu \nu}(t) = \hat{T}_{\mu \nu}(t_0)+ \int \Delta^{\alpha \beta}  \left[ \hat{T}_{\mu \alpha} \hat{T}_{\beta \nu} \right] + \int \frac{1}{2}d\Sigma_\mu \beta_\nu
  \hat{\Pi}_{\alpha \beta} \partial^\alpha \beta^\beta 
\end{equation}
the first is a Brownian integral, over ``kicks'' $\Delta^{\alpha \beta}$, where in the Ward identity always fixes one component (Eq. \ref{ward} integrated by parts).    The second is an integral over time.   In the Gaussian limit these integrals always converge, and Eq. \ref{gaussian} plays a role of the fluctuation-dissipation relation.

Note that Eq. \ref{itoeq} includes in it ideal evolution, where, because of the fact that no entropy is created the first term is certain (probability unity) and the second term is zero (Eq. \ref{deltasid}).   Stochastic steps, however, sample over both thermal fluctuations and dissipative evolution.  In a highly turbulent regime (where $\zeta^\mu \sim u^\mu$), it will be ``likely'' that thermal fluctuations will bring the system far away from equilibrium.

Let us try to sketch how to implement these equations in a solution.   Eq. \ref{t0def} and Eq. \ref{pidef} could in principle be used to define, perhaps on a discrete lattice, a $\beta_\mu$ field and $n_\mu$ foliation out of any energy-momentum tensor at a given time $t$.  After integrating Eq. \ref{itoeq} with a Metropolis procedure based around Eq. \ref{crookszeta} one can 
\begin{equation}
  \label{zrec}
  \left. \lnz\right|_{t+dt} =  \int\left. \mathcal{D}g_{\mu \nu}(x) T^{\mu \nu}\right|_{t+dt}  \eqcomma \left. \beta_\mu \right|_{t+dt} = \frac{\left. \delta \lnz\right|_{t+dt}}{\delta T_{\mu \nu}} n_\nu 
    \end{equation}
equation \ref{pidef} and \ref{t0def} can then be used to orient $T_0^{\mu \nu}$ and $\Pi^{\mu \nu}$ at the new step, and the Ward identity Eq. \ref{ward} can be used for relative normalization.  The cycle would then restart.   Such a procedure, using metropolis-type sampling at each time-step, would be computer-intensive but achievable and would be a logical sequel to the static lattice fluctuation study examined in \cite{burch}.
Afterwards, the observable $T_{\mu \nu}$ correlators can be sampled the usual way numerically
\[\  \ave{\prod_i T^{\mu_i \nu_i}(x^{\gamma_i})} = \ave{\prod_i \left(T^{\mu_i \nu_i }_0 (x^{\gamma_i})+\Pi^{\mu_i \nu_i } (x^{\gamma_i}) \right) }     \]
The ingredients input from microscopic theory for such a simulation are the entropy content of a configuration of $\beta_\mu$ in each cell, equation \ref{pidef} and \ref{t0def} and the short-range structure of the commutator at thermal equilibrium
\begin{equation}
  \label{Gmunu}
  G_{\mu \nu \alpha \beta}(x-x',T) =\ave{ \left. \left[ \hat{T}_{\mu \nu}(x'),\hat{T}_{\alpha \beta}(x) \right]} \right|_{(x-x')_\mu/\beta^\mu\ll 1}
  \end{equation}
to close the Ward identity Eq. \ref{wardpart} or calculate the statistical distribution of $\int \hat{T}_{\mu \nu}dx^\mu$.  The ingredients are therefore exactly the same as those required for to solve a general viscous hydrodynamics, where terms of the Taylor expansion of the Fourier transform of Eq. \ref{Gmunu} are matched to the gradient order.  However, as mentioned earlier, the advantage of this approach is that thermal fluctuations should be resummed at each step.

Let us close this section with some qualitative considerations of what such stochastic dynamics will look like.

If the energy-momentum tensor is isotropic in the co-moving frame with $\beta_\mu$, Eq. \ref{alphabeta} can be used to put \ref{pidef} to zero, what is sometimes called ''hydrodinamization''.
   That said, even in this case, the RHS of Eq. \ref{zgaussian} is not zero.
   This means that $T^{\mu \nu}$ at time $t+dt$ will generally be different from that predicted by ideal hydrodynamic flow without fluctuations.   This illustrates how, if Crooks fluctuation theorem is assumed, dissipation within a hydrodynamic evolution arises inevitably, driven by microscopic fluctuations, as discussed in \cite{mooresound}.

   If turbulence and fluctuations do not dominate, what is the average limit of this stochastic evolution?   If Jumps are determined by Eq. \ref{hydrocrook} it is clear that ``over many steps'' a global maximization of entropy is reached, i.e. the system always tends towards the maximum entropy state, as required.
   Furthermore the ward identify Eq. \ref{wardpart} will ensure the conservation of the average momentum current
   \begin{equation}
     \label{conslaw}
     \label{maxcat}
\partial_\mu \ave{\hat{T}^{\mu \nu}}=0 \eqcomma  \partial_\mu \ave{\hat{T}^{\mu \nu}_0} = - \partial_\mu \ave{\hat{\Pi}^{\mu \nu} }
     \end{equation}
    Integrating by parts the second term of Eq. \ref{itoeq} over a time scale of many $\Delta_{\mu \nu}$ 
   gives, in a frame comoving with $d\Sigma_\mu$
   \begin{equation}
     \int_{0}^\tau d\tau' \ave{\hat{\Pi}_{\mu \nu}} \partial^\mu \beta^\nu \sim \beta^\mu \partial_\mu \ave{\hat{ \Pi}_{\mu \nu}} + \ave{\hat{\Pi}_{\mu \nu}} = F(\partial^{n\geq 1} \beta_\mu,...)
   \end{equation}
   where $F(\beta_\mu)$ is independent of $\Pi_{\mu \nu}$.   Because local entropy is maximized by Eq. \ref{rhodef}, $F()$ cannot depend on $\beta_\mu$ but only on gradients.
   Because of the $it\hat{o}$ isometry, the long term expectation values of Eq. \ref{itoeq} are equal in squares
   \begin{equation}
     \label{israeq}
\int \Delta^{\alpha \beta} \ave{\left[ T_{\mu \alpha} T_{\beta \nu}  \right]}^2 =  \int \frac{1}{4} d\Sigma_\mu\ave{ \beta_\nu  \Pi_{\alpha \beta} \partial^\alpha \beta^\beta }^2
     \end{equation}
   Hence, the equation of motion for the long-time average of $\Pi_{\mu \nu}$ should approach Israel-Stewart type dynamics used earlier \cite{kodama} when fluctuations are neglected.
   
   Therefore, equation \ref{zrec} allows us to do one better,
   reconstructing the partition function and the field of Lagrange multiplies $\beta_\mu$ at each point in time.
   The simplest, Gaussian approximation, together with the fluctuation-dissipation relation would mean
   \begin{equation}
\ave{\left(\Delta \hat{\Pi}_{\mu \nu}(x) \right)^2} \sim \tau_\pi
   \end{equation}
   which already is beyond reach of an expansion such as Eq. \ref{sk}.
   However, we can do better still:  Zubarev/Crooks dynamics allows to sample probability distributions of every observable, including higher cumulants.
   We expect such effects will be important in the deep turbulent regime or close to the critical point \cite{stephcrit}.   

In this respect it is worth mentioning that
recently, an effort to construct first order stable theories \cite{kovtun1st,gavassino,shokri} has provided indications that theories written to first-order in gradient are stable, provided one allows for ``off-shell'' small violations of the second law of hydrodynamics.   Such ``off-shell fluctuations'' (``off-shell'' means not obeying the equation of motion), for stability, require that entropy is bounded \cite{gavassino,shokri}.
Our theory could unify this picture with the more traditional Israel-Stewart approach as we describe how violations of the second law of thermodynamics occur in the ``fluctuating'' part of the energy-momentum tensor, related to the dissipative part via Crook's theorem, where it is clear that bounded fluctuations around the average (and the smallness of the dissipative term in Eq. \ref{kmscond}) require a bounded entropy.   We therefore speculate that the second order term in Eq. \ref{itoeq} will lead to something like Israel-Stewart, and the first term, averaged over long times, will give something like \cite{kovtun1st}.
For a series of Crook's steps to fluctuate around a deterministic equation (the It$\hat{o}$ isometry to be satisfied), one needs entropy to be bounded (if not, deviations will fluctuate to infinity), thereby confirming the intuition of \cite{gavassino,shokri}.

Finally, we remark that as discussed in \cite{landi,landizub}, a remarkable range of systems in principle well away from any kind of hydrodynamic limit (nano-engines, folding proteins and so on) seem to saturate thermodynamic uncertainity relations.   A qualitative explanation is that thermal fluctuations together with the chaotic regime {\em help} in the fast equilibration of the system.   Our hope is that this dynamics, in the turbulent regime, could result in large hydrodynamic fluctuations in small systems {\em helping} achieving thermalization in such systems, as seen in \cite{cms}.
\section{Discussion and conclusions}
The formalism developed here could be straight-forwardly extended for more complicated microscopic theories according to the prescriptions outlined in \cite{zubarev}, via the substitution, in the exponent of Eq. \ref{rhodef} \cite{palermo}
\begin{equation}
  \beta_\mu T^{\mu \nu} \rightarrow  \beta_\mu T^{\mu \nu} + \mu J^\mu + \mathcal{W} \mathcal{J}^{\mu }
  \end{equation}
where $\mu$ is the chemical potential for a conserved charge and $J^\mu$ is that conserved charge's current \cite{becstat}, $\mathcal{W}$ is the vortical susceptibility and $\mathcal{J}_{\mu}$ the angular momentum (note that as as shown in \cite{gt3} this term will, necessarily for causality, be augmented by a relaxation timescale.  \cite{zubpol} and \cite{ghosts} also show care needs to be taken with gauge symmetries).
The form of the Crooks relation Eq. \ref{hydroloop}, and subsequent formulae should not change, since this non-equilibrium definition of entropy is universal.
What changes is that $\Pi_{\mu \nu}$ will get contributions from charge conductivity, polarization currents (allowing a derivation of casual magnon dissipation \cite{gt3}).   Even gauge currents \cite{ghosts} can be accommodated by adding the pure-gauge current $U_a \partial_\mu U_b$, making all exponents gauge covariant and tracing over the color fields.
In all of these cases, the fluctuation and dissipation evolution can be integrated in a similar manner.

This approach is likely to resolve some paradoxes that arises from the construction of the ``equilibrium'' state of \cite{palermo,prokhorov}.   The authors of these works have constructed the ``equilibrium'' Wigner function and density matrix for a free gas prepared to be in an arbitrary, accellerating and rotating reference frame (A ``passive'' coordinate transformation).
Equilibrium is enforced through the maximization of entropy, via Lagrange multipliers, in the laboratory frame.   However, it can be easily checked that the KMS condition does not hold.   How can an evolving state be in equilibrium, and vice-versa?    The answer is, of course, that entropy is a frame-dependent quantity and the KMS state is defined via a local ``co-moving'' frame, where the second law of hydrodynamics also holds.   The non-inertiality will generally contain energy that can be transformed into work, and thus ``global equilibrium'' w.r.t. that frame does not preclude further increases in entropy.   The current approach, provided the system is close to {\em local} equilibrium, allows us to calculate it's further dynamics in a way that converges to the local fulfillment of KMS conditions independently of the foliation.

The operator representation of the hydrodynamic $\hat{T}_{\mu \nu}$ and $\hat{\Pi}_{\mu \nu}$ in terms of density matrices also invites investigation weather the dynamics developed here can be linked directly to the dynamics of the reduced density matrix of the multi-particle system via the Hamiltonian flow \cite{zhang,reyes,casini}.  While entanglement between $\hat{T}_{\mu \nu}$ and $\hat{\Pi}_{\mu \nu}$ is neglected in the dynamics, this approach can link to the entanglement entropy via relative normalization, end hence give an insight into the quantum entanglement between microscopic and macroscopic degrees of freedom.   Given that entanglement of the QCD initial state in heavy ion collisions is actively being investigated \cite{kharzeev}, this might lead to a phenomenology of our approach.

Finally, as an extremely speculative application of this formalism takes inspiration from Analogue gravity, where hydrodynamics has long been known to describe the kinematical \cite{analogue}, rather than dynamical part of general relativity., general relativity can be derived \cite{jacobson} as an ``equation of state'' with entropy and horizons being related as an assumption.      Perhaps including a horizon term (as \cite{jacobson} does locally in terms of the Congruence) in the RHS of equation \ref{hydroloop} would lead to a well-defined fluctuation-dissipative dynamics involving a spacetime obeying exact diffeomorphism invariance \cite{holo}:  On a basic level, random fluctuations mean ``one initial condition leads to many final states'', while dissipation implies the opposite, ``many initial conditions to one final state''.  A non-unitary dynamics where probabilistic fluctuations and dissipation are correlated can keep track of quantum corrections to gravity (such as horizon entropy) are included but all correlators are generally covariant. 
If nothing else, a fundamentally theory that ``the universe is governed by Crooks'' would have a considerable observational support!

In conclusion, we used the Crooks fluctuation theorem and Zubarev hydrodynamics to construct an equation of motion for fluctuating hydrodynamics based on the evolution of the partition function.

Given an initial ensemble of $\hat{T}_{\mu \nu}(\Sigma_\mu)$, where $\Sigma_\mu$ is a foliation, provided there is a $\beta_\mu (\Sigma_\mu)$ field whose Zubarev partition function Eq. \ref{kmscond} reproduces $\hat{T}_{\mu \nu}$ to a good approximation, this paper proposes a stochastic procedure to evolve this ensemble so that it maintains approximate local equilibrium, with the deviation from equilibrium also being counted as an ensemble $\hat{\Pi}_{\mu \nu}$.
Our results should converge to the usual limits (Israel Stewart hydrodynamics, Boltzmann statistical mechanics) in the right limits (respectively small fluctuations and small gradients), 
but be fluctuation-dominated for smaller systems.     We speculate the interplay between fluctuations and non-linearity could help small ``dollops of fluid'' seen in data \cite{cms} equilibrate quickly  and
hope to develop an analytical and  numerical phenomenology, the latter on a lattice, for this approach in the future.

{\bf Acknowledgements}
This work was was made possible by Francesco Becattini's hospitality in INFN Firenze, as well as crucial suggestions, discussions and criticisms.\\
We would like to thank Sangyong Jeon, Gabriel Landi, Giacomo Guarnieri and Massoud Shokri for discussions.\\
 GT acknowledges support from FAPESP proc. 2017/06508-7,
participation in FAPESP tematico 2017/05685-2 and CNPQ bolsa de
 produtividade 301432/2017-1.
\appendix*
\section{Isolating the equilibrium part of the density matrix \label{appendixz}}
Here we shall clarify what we mean in Eq. \ref{piTdef}, and how Eq. \ref{kmscond} is developed.
  The definition of what $\hat{T_0}^{\mu \nu}$ is can be deduced from the definition of the equilibrium density matrix $\hat{\rho}_T$ in terms of the Hamiltonian (the integral over a spacetime foliation of the Hamiltonian density for a field)
\begin{equation}
  \label{densmatdef}
\hat{\rho}_{T} = \frac{1}{\mathcal{Z}}\exp \left[- \frac{\hat{H}}{T}\right] \eqcomma \ave{\hat{O}}=\mathrm{Tr} \left( \hat{O}\hat{\rho}_T \right) = \frac{\delta}{\delta J_O} \lnz
\end{equation}
$J_O$ is a general ``source'', used to get expectation values (for energy-momentum tensors, the metric in a certain frame can be used).

For a quantum field, given any microscopic Lagrangian density $L$, a flow field $\beta_\mu(x,t)$ and a foliation $d\Sigma_\mu$, equation \ref{densmatdef} can be generalized into a density matrix describing a system {\em with that microscopic lagrangian} prepared to be {\em instantaneously} in local equilibrium.    That density matrix is given by Eq. \ref{rhodef}, in terms of a stress-energy tensor we call $T_0^{\mu \nu}$.    Zubarev has proved (see \cite{zubarev} and references therein) that if an instantaneus foliation Eq. \ref{foldef} is found where the system is in perfect local equilibrium subsequent quantum evolution of the system is given by Eq. \ref{rhodef}.    Of course such a carefully prepared state is generally impossible, and in this work it must only be true {\em approximately} (see Footnote 3).

One can then use Eq. \ref{rhodef} in conjunction with Eq. \ref{densmatdef} to calculate the probability to get any cumulant of the energy momentum tensor
\begin{equation}
\label{t0eta}
  \ave{\Delta  T^{\mu \nu}_0(x_1,t_1)... \Delta T^{\mu \nu}_0(x_n,t_n)} =\ave{ \frac{\delta^n}{\delta \eta'_{\mu \nu}(x_1,t_1) .... \eta'_{ \mu \nu}(x_n,t_n)} \lnz}
\end{equation}
$\eta'_{ \mu \nu}$ is the metrix which is, at a given $x,t$, at rest w.r.t. $\beta_\mu(x,t)$.  This is not generally an inertial frame, and equation Eq. \ref{t0eta} is not to be confused with the general definition of the energy momentum tensor w.r.t. metric Eq. \ref{zpartdiss}.

Since the system is however never usually in full local equilibrium, Eq. \ref{rhodef} does not commute with $\hat{H}$.  Hence, $T^{\mu \nu}_0$ is usually not the full $T^{\mu \nu}$ (it might be at a given instant,but not before or after), and Equation \ref{rhodef} has no information about $T^{\mu \nu}-T^{\mu \nu}_0$.

We remember equation \ref{ergodic}, so observables specific of each volume cell is described by some $\hat{P}(\mu)$, determined by an average over states.
We shall implement Eq. \ref{piTdef} by assuming that the probability functional $\mathcal{P}(...)$ for a moment in time\footnote{For a general probability distribution function $\mathcal{P}(X)$ $\lnz$ is the cumulant generating function 
\[\   
\lnz =  \left. \ln \int \mathcal{P}(X) \exp [tX] dX  \right|_{t=1}  \]
The inverse with the Boltzmann factor gives the usual partition function of statistical mechanics
}
of the total energy momentum tensor (defined as in Eq. ) factorizes classically into 
\begin{equation}
  \label{factp}
\mathcal{P}(T^{\mu \nu},t) = \int  \mathcal{P}(T_0^{\mu \nu},t) \mathcal{P} (\Pi_{\mu \nu},t) \delta\left( T^{\mu \nu} - T_0^{\mu \nu} - \Pi^{\mu \nu} \right)  \mathcal{D} T_0^{\mu \nu} \mathcal{D} \Pi^{\mu \nu}
\end{equation}
where the cumulants of $T_0^{\mu \nu}$ are given by Eq. \ref{rhodef} and \ref{t0eta} and  $\mathcal{P} (\Pi_{\mu \nu})$ is something we have to find out.

Note that as remarked in section \ref{eqsection}, the division in $T_0^{\mu \nu}$ and $\Pi_{\mu \nu}$ is not unique, analogously to the ``Hamiltonian of weak force'' decomposition \cite{hamweak1,hamweak2} and only the sum remains observable\footnote{In the Quantum mechanics of Eq. \ref{zuberdef} with one coordinate $x$ the equivalent procedure would be to construct $x=y+z$ chosen so that \[\ \qave{y|H|y'}\ll \qave{y|\Sigma|y'} \eqcomma \qave{z|\Sigma|z'}\ll \qave{z|H|z'} \eqcomma \qave{y|H|z},\qave{y|\Sigma|z} \ll 1 \] and then expand.   In systems with a coupling to a bath close to the Markovian limit this choice is generally possible, but its uniqueness and observability is controversial \cite{hamweak1,hamweak2}}

Also, note that Eq. \ref{factp} does not preclude correlations, just quantum entanglement, between the two components.  In other words, in analogy with Eq. \ref{zuberdef}, $[ \hat{T}_{\mu \nu},\hat{\Pi}_{\mu \nu}]$ is negligible, either because it is Knudsen-suppressed or because of decoherence.  This commutator should go as $\sim \alpha K$ and higher in Eq. \ref{hyerarchyscales}.

We shall further hope that in the strongly interacting theory $\hat{T}_0^{\mu \nu}$ is ``close'' to $\hat{T}^{\mu \nu}$, in the hope of using Crooks fluctuation theorem to calculate ``the rest'' (as \cite{landizub} did for quantum mechanics).

To implement the above quantitatively,  we take the definition of the density matrix in \cite{nishioka} in the configuration space basis. Ignoring the problems of normalization in QFT, we get that to construct a density matrix one needs the microscopic Lagrangian and the asymptotic conditions of the field configurations (here, $y(0^\pm)$.  Note that in 0+1D Quantum Mechanics all choices are equivalent because of the Stone Von Neumann theorem, but in higher dimensional quantum field theory they are not).    In a configuration space basis 
\begin{equation}
\label{nishioka}
        \langle x| \,\rho\, |x' \rangle = \frac{1}{\mathcal{Z}} 
\int^{\tau=\infty}_{\tau=-\infty} \int \left[ \mathcal{D}\phi, \mathcal{D} y(\tau)\,\mathcal{D} y'(\tau)\right]\,e^{- iS (\phi y, y')}
        \cdot \delta \left[ y(0^+) - x'\right]\,\delta \left[ y'(0^-) - x\right]\ ,
\end{equation}
where $\tau$ is the proper time and $0^{\pm}$ refers to the asymptotics in the foliation defined by $\Sigma_\mu$.   Following  \cite{peskin}
\[\   \delta \left[ y(0^+) - x'\right]\,\delta \left[ y'(0^-) - x\right] = \frac{\delta J_i(y(0^+))}{\delta J_i(x')}  \frac{\delta J_j(y(0^-))}{\delta J_j( x)}  \]
Hence, by integrating by parts the density matrix can be obtained from a partition function 
\begin{equation}
  \label{denspart}
  \langle x| \,\rho\, |x' \rangle  = \frac{\delta^2 }{\delta J_i(x) \delta J_j (x')} \lnz(J_i(y(0^+)+J_j(y'(0^-))
\end{equation}
We must remember that in Quantum mechanics $Z$ is a function and $\hat{\rho}$ a matrix of numbers, in Quantum field theory $Z$ is a functional and $\hat{\rho}$ a ``matrix of functions'', generated by an appropriate choice of $J_{i,j}(x,t)$.

If the system is close to local equilibrium, the Matsubara technique can be used \cite{kapustagale} to fix $y(0^\pm)$:   One can construct a finite temperature partition function by imposing, on functional integrals of fields $\phi$ (the microscopic DoFs here), the condition that $\phi(x,t)=\pm \phi(x,t+i/T)$ ($\pm$ refers to spin-statistics).
Given a choice of a flow field $\beta_\mu$ and an instant $\Sigma_\mu$ of a foliation an equilibrium partition function can be computed in the reference frame at rest w.r.t. $\beta_\mu(x_i^\mu)$ \cite{gale}
\begin{equation}
  \label{tempz}
  Z_{T_0}(J(y))=  \int \mathcal{D} \phi \exp \left[ - \int_0^{T^{-1}(x_i^\mu)} d\tau' \int d^3 x \left( L(\phi) +J(y)\phi\right)  \right]
\end{equation}
The periodicity of the time integral in Eq. \ref{tempz} mean that an arbitrary $J_i(x^\mu) \rightarrow J_i\left(T^{-1}\left(\vec{x},t\right),\vec{x}\right)$ in the rest frame where $\beta_\mu=(T^{-1},\vec{0})$.  This ensures that 
Eq. \ref{kmscond}  follows.

Just like $\hat{T}^{\mu \nu} \ne \hat{T}_0^{\mu \nu}$, the $Z$ for the microscopic theory is not equal to $Z_{T_0}$.   However, for any $\beta_\mu,d\Sigma_\mu$ at that instant we can factorize $Z=Z_{T_0} \times Z_\Pi$, where $Z_\Pi$ is simply ``the rest''. 
Note that the normalization of Eq. \ref{kmscond} is taken care automatically because of the definition of $\hat{\rho}$ in terms of $Z$:  An expansion of the form $Z=Z_{T_0}\times Z_{\Pi}$ with $Z_{\Pi}$ would automatically result in Eq \ref{kmscond}, with the correct normalization, holding.

The above procedure is possible {\em always}, for any choice $\beta^\mu(x,t)$ and any $\hat{T}_{\mu \nu}$.
For this paper's results to be applicable, however, one needs that, given a 
given a partition function ``localized in time'', $J_{i,j}\propto \delta(\tau-\tau')$, the partition function at the next time step  $J_{i,j}\propto \delta(\tau-\tau'+\Delta )$ is given through Eq. \ref{hydrocrook}.   This allows us to use Eqs. \ref{zgaussian},\ref{wardpart} to reconstruct the partition function $Z$ at {\em all} times.

When is this true, if ever?   To answer this question, we recall that the usual hydrodynamics derivation depends on the assumption that each particle's mean free path is large enough that it defines a volume in the termodynamic limit, but small enough w.r.t. the gradients of the fluid.  This is equivalent to saying that the ``infinities'' in the integral in Kubo's formula \cite{kadanoff} are still ``small'' w.r.t. the hydrodynamic gradients.

In our case, something similar happens. Eq. \ref{nishioka} is defined in terms of asymptotic limits, $0^\pm$.  If each $d\Sigma_\mu$ is ``long'' w.r.t. some microscopic scale, approximating
\begin{equation}
  \label{hyerarchy}
 \beta_\mu d\Sigma^\mu \equiv \Delta \simeq 0^+ - 0^-
  \end{equation}
allows us to compare Eq. \ref{tempz} to Eq. \ref{nishioka} at a given time step so that the two can be approximately the same.

There is however a further issue:
   $Z_{T_0}$ and $Z_\Pi$ are not generally independent, since $\hat{T}_0$ and $\hat{\Pi}$ are generally correlated.  However, relations such as Eq. \ref{zuberdef} are valid if $\hat{\Pi}$ and $\hat{T_0^{\mu \nu}}$  commute, an assumption equivalent to the Markovian fluctuations assumption needed to derive Crooks's theorem.

In \cite{landizub} the commutativity is manifest by the fact that $\hat{H}$ and $\hat{N}$ in equation \ref{zuberdef} is stationary.   In our case, Eq. \ref{hyerarchy} means that commutativity between $\hat{T}_0^{\mu \nu}$ and $\hat{\Pi}$ will be of order $>\Delta$.   One then see explicitly, within the modular Hamiltonian representation, that the infinite tower of nested correlators \cite{zhang} will correspond to an expansion in powers of $\Delta$.    Using Crooks fluctuation theorem then means $\left(\omega_{\mu  \nu} \omega^{\mu \nu}\right)^{1/2}$ in Eq. \ref{hydrocrook} is ``large'' w.r.t. $\Delta$.   According to Eq. \ref{hyerarchyscales} $\Delta\sim \order{\alpha K/(\partial_\mu u_\nu)}$

In summary, the separation of scales for the applicability of our results looks very much related to the applicability of usual hydrodynamics.  What our approach has, as an advantage over the usual approach of considering only equations of motion for the averages, is that fluctuation and dissipation are treated on the same footing via operators.   This means that while expanding in Knudsen number we ``keep all thermodynamic fluctuations'', i.e. the expansion in te microscopic length-scale'' (equation (1) of \cite{ryblewski} and \cite{tinti}) into account.
As an analogy, one could consider coarse-graining Quantum Chromodynamics in terms of Wilson loops rather than in terms of chiral perturbation theory.   This, in principle, allows us to keep fluctuations at sub-hadronic scale, at the price of not having manifest hadronic degrees of freedom.

We should also take a moment to compare the above derivation with that of \cite{zubarev}:  The authors of \cite{zubarev} start with the Von Neumann definition of entropy and expand it around a foliation, defining $\beta_{\mu}$ (and chemical potentials) as a field of Lagrange multipliers.   This is appropriate if the system, and every point spanned by $n_\mu$ is close to global equilibrium,so deviations from maximization of entropy are small and the entropy in each volume element foliated by $n_\mu$ is not too far from the maximum.   In contrast, building $\hat{T}^{\mu \nu}$ locally by deriving most of it from the KMS condition is appropriate if every point of the system is close to {\em local} equilibrium, irrespective of how far away we are from global equilibrium.   This is not a trivial consideration, since, if one considers the hydrostatic limit with a small perturbation and a viscosity $\eta$, the local equilibration timescale ($\sim \eta/(Ts)$ is inversely proportional to the global one $\sim s/(\eta k)$ where $k$ is the wavenumber of the sound-wave).  The EFT defined here and the one in \cite{zubarev} have opposite domains of validity, although their zero-th order equilibrium terms are the same. 


\begin{thebibliography}{10}

\bibitem{crooks} G. Crooks, cond-mat/9901352\\
  \url{https://en.wikipedia.org/wiki/Crooks_fluctuation_theorem}

  
\bibitem{zubarev} 
  F.~Becattini, M.~Buzzegoli and E.~Grossi,
  Particles {\bf 2}, no. 2, 197 (2019)
  doi:10.3390/particles2020014
  [arXiv:1902.01089 [cond-mat.stat-mech]].


  
\bibitem{kodama}
R.~Derradi de Souza, T.~Koide and T.~Kodama,
Prog. Part. Nucl. Phys. \textbf{86} (2016), 35-85
doi:10.1016/j.ppnp.2015.09.002
[arXiv:1506.03863 [nucl-th]].


\bibitem{cms}
  V.~Khachatryan {\it et al.} [CMS Collaboration],
  Phys.\ Lett.\ B {\bf 765}, 193 (2017)
  doi:10.1016/j.physletb.2016.12.009
  [arXiv:1606.06198 [nucl-ex]].


  \bibitem{landau} E.M.~Lifshitz, L.D.~Landau, \textit{Fluid Mechanics},
Butterworth-Heinemann, 1987, (Russian original: \textit{Gidrodinamika},
State Publishing House for Physics-Mathematics Literature, Moscow
 L.Lifshitz and Landau and Pitaevski, Statistical mechanics part 2 ( Volume 9)

  \bibitem{kovtun}
P.~Kovtun,
J. Phys. A \textbf{45}, 473001 (2012)
doi:10.1088/1751-8113/45/47/473001
[arXiv:1205.5040 [hep-th]].

\bibitem{stephanov}
X.~An, G.~Basar, M.~Stephanov and H.~Yee,
Phys. Rev. C \textbf{100}, no.2, 024910 (2019)
doi:10.1103/PhysRevC.100.024910
[arXiv:1902.09517 [hep-th]].


\bibitem{stephcrit}
  M.~Bluhm, M.~Nahrgang, A.~Kalweit, M.~Arslandok, P.~Braun-Munzinger, S.~Floerchinger, E.~S.~Fraga, M.~Gazdzicki, C.~Hartnack, C.~Herold, R.~Holzmann, I.~Karpenko, M.~Kitazawa, V.~Koch, S.~Leupold, A.~Mazeliauskas, B.~Mohanty, A.~Ohlson, D.~Oliinychenko, J.~M.~Pawlowski, C.~Plumberg, G.~W.~Ridgway, T.~Schäfer, I.~Selyuzhenkov, J.~Stachel, M.~Stephanov, D.~Teaney, N.~Touroux, V.~Vovchenko and N.~Wink,
[arXiv:2001.08831 [nucl-th]].

\bibitem{gale}
M.~Singh, C.~Shen, S.~McDonald, S.~Jeon and C.~Gale,
Nucl. Phys. A \textbf{982}, 319-322 (2019)
doi:10.1016/j.nuclphysa.2018.10.061
[arXiv:1807.05451 [nucl-th]].

\bibitem{csernai}
Z.~Lazar, L.~Csernai, D.~Molnar and I.~Lazar,
``Fluctuation and dissipation in discretized fluid dynamics,''\\
Correlations and Fluctuations '98 (CF 98), 539-546\\
L.~Csernai, S.~Jeon and J.~I.~Kapusta,
Phys. Rev. A \textbf{56}, 6668 (1997)
doi:10.1103/PhysRevE.56.6668
[arXiv:nucl-th/9708033 [nucl-th]].

\bibitem{ryblewski}
D.~Montenegro, R.~Ryblewski and G.~Torrieri,
Acta Phys. Polon. B \textbf{50}, 1275 (2019)
doi:10.5506/APhysPolB.50.1275
[arXiv:1903.08729 [hep-th]].

\bibitem{giorgio} 
  G.~Torrieri,
  Phys.\ Rev.\ D {\bf 85}, 065006 (2012)
  doi:10.1103/PhysRevD.85.065006
  [arXiv:1112.4086 [hep-th]].

  \bibitem{burch}
T.~Burch and G.~Torrieri,
Phys. Rev. D \textbf{92}, no.1, 016009 (2015)
doi:10.1103/PhysRevD.92.016009
[arXiv:1502.05421 [hep-lat]].

\bibitem{glorioso}
H.~Liu and P.~Glorioso,
PoS \textbf{TASI2017}, 008 (2018)
doi:10.22323/1.305.0008
[arXiv:1805.09331 [hep-th]].

\bibitem{grozdanov}
S.~Grozdanov and J.~Polonyi,
Phys. Rev. D \textbf{91}, no.10, 105031 (2015)
doi:10.1103/PhysRevD.91.105031
[arXiv:1305.3670 [hep-th]].

  \bibitem{lagrangian} 
  D.~Montenegro and G.~Torrieri,
  Phys.\ Rev.\ D {\bf 94}, no. 6, 065042 (2016)
  doi:10.1103/PhysRevD.94.065042
  [arXiv:1604.05291 [hep-th]].

  
    \bibitem{nicolis}
  S.~Endlich, A.~Nicolis, R.~Rattazzi and J.~Wang,
  JHEP {\bf 1104}, 102 (2011)
  [arXiv:1011.6396 [hep-th]].


\bibitem{gripaios}
  B.~Gripaios and D.~Sutherland,
  Phys.\ Rev.\ Lett.\  {\bf 114}, no. 7, 071601 (2015)
  doi:10.1103/PhysRevLett.114.071601
  [arXiv:1406.4422 [hep-th]].


  \bibitem{tinti} 
D.~Montenegro, L.~Tinti and G.~Torrieri,
Phys. Rev. D \textbf{96}, no.5, 056012 (2017)
doi:10.1103/PhysRevD.96.056012
[arXiv:1701.08263 [hep-th]].


\bibitem{gt3}
  D.~Montenegro and G.~Torrieri,
  Phys.\ Rev.\ D {\bf 100}, no. 5, 056011 (2019)
  doi:10.1103/PhysRevD.100.056011
  [arXiv:1807.02796 [hep-th]].

\bibitem{linear}
D.~Montenegro and G.~Torrieri,
[arXiv:2004.10195 [hep-th]].




\bibitem{becborn}
F.~Becattini,
[arXiv:0901.3643 [hep-ph]].

\bibitem{hartnoll}
L.~V.~Delacrétaz, T.~Hartman, S.~A.~Hartnoll and A.~Lewkowycz,
JHEP \textbf{10}, 028 (2018)
doi:10.1007/JHEP10(2018)028
[arXiv:1805.04194 [hep-th]].

\bibitem{kms}
R.~Haag, N.~Hugenholtz and M.~Winnink,
Commun. Math. Phys. \textbf{5}, 215-236 (1967)
doi:10.1007/BF01646342

\bibitem{kapustagale}
J.~Kapusta and C.~Gale,
``Finite-temperature field theory: Principles and applications,''
doi:10.1017/CBO9780511535130

\bibitem{nishioka}
  T.~Nishioka,
  Rev.\ Mod.\ Phys.\  {\bf 90}, no. 3, 035007 (2018)
  doi:10.1103/RevModPhys.90.035007
  [arXiv:1801.10352 [hep-th]].

\bibitem{wild} Camillo De Lellis, Laszlo Szekelyhidi Jr, Arch. Ration. Mech. Anal. 195 (2010), no. 1, 225-260, 0712.3288

\bibitem{vicol} Tristan Buckmaster and Vlad Vicol, Annals of Mathematics
  Vol. {\bf 189} , No. 1 (January 2019), pp. 101-144  arXiv:1709.10033

  \bibitem{vicol2} Tristan Buckmaster and Vlad Vicol, arXiv:1901.09023 

 \bibitem{vicol2d} Peter Constantin, Andrei Tarfuleia and Vlad Vicol, arXiv:1305.7089   

 \bibitem{zeroth}
B.~R.~Pearson, T.~A.~Yousef, N.~E.~L.~Haugen, A.~Brandenburg and P.~A.~Krogstad,
Phys. Rev. E \textbf{70}, 056301 (2004)
doi:10.1103/PhysRevE.70.056301
[arXiv:physics/0404114 [physics]].
    
\bibitem{peskin}  Peskin and Schroeder, an introduction to Quantum field theory
  
    
\bibitem{landi} 
Andre M. Timpanaro, Giacomo Guarnieri, John Goold, Gabriel T. Landi, 1904.07574


\bibitem{landizub} Giacomo Guarnieri, Gabriel T. Landi, Stephen R. Clark, John Goold,Phys. Rev. Research {\bf 1}, 033021 (2019)
 [arXiv:1901.10428  [quant-ph]]

 
\bibitem{bass} R.Bass, ``Stochastic processes'', Cambridge University Press (2011)
 
 \bibitem{zhang}
Z.~Jiang,
Phys. Rev. A \textbf{89}, no.3, 032128 (2014)
doi:10.1103/PhysRevA.89.032128
[arXiv:1310.2687 [quant-ph]].

\bibitem{hamweak1} P. Talkner and P. Hanggi {\it Colloquium:} Statistical Mechanics and Thermodynamics at Strong Coupling: Quantum and Classical, 
[arXiv:1911.11660 [quant-ph]].

\bibitem{hamweak2} P. Strasberg,M. Esposito, Phys. Rev. E \textbf{101}, 050101 (2020)
[arXiv:2001.08917 [quant-ph]].

  
  \bibitem{zubpol}
F.~Becattini,
[arXiv:2004.04050 [hep-th]].


\bibitem{palermo}
F.~Becattini, M.~Buzzegoli, A.~Palermo and G.~Prokhorov,
[arXiv:2009.13449 [hep-ph]].

\bibitem{prokhorov}
G.~Y.~Prokhorov, O.~V.~Teryaev and V.~I.~Zakharov,
JHEP \textbf{03} (2020), 137
doi:10.1007/JHEP03(2020)137
[arXiv:1911.04545 [hep-th]].

\bibitem{becstat}
F.~Becattini,
Phys. Rev. Lett. \textbf{108}, 244502 (2012)
doi:10.1103/PhysRevLett.108.244502
[arXiv:1201.5278 [gr-qc]].

  

\bibitem{romentropy} 
  P.~Romatschke,
  Class.\ Quant.\ Grav.\  {\bf 27}, 025006 (2010)
  doi:10.1088/0264-9381/27/2/025006
  [arXiv:0906.4787 [hep-th]].


    \bibitem{kadanoff} L.Kadanoff and P.Martin, Annals of Physics {\bf} 24 419-469 (1963)

  
  

\bibitem{boulware}
  S.~Deser and D.~Boulware,
  J.\ Math.\ Phys.\  {\bf 8}, 1468 (1967).
  doi:10.1063/1.1705368

\bibitem{jeon} 
  A.~Czajka and S.~Jeon,
  Phys.\ Rev.\ C {\bf 95}, no. 6, 064906 (2017)
  doi:10.1103/PhysRevC.95.064906
  [arXiv:1701.07580 [nucl-th]].

    \bibitem{jarz}
M.~Caselle, G.~Costagliola, A.~Nada, M.~Panero and A.~Toniato,
Phys. Rev. D \textbf{94}, no.3, 034503 (2016)
doi:10.1103/PhysRevD.94.034503
[arXiv:1604.05544 [hep-lat]].
  
  \bibitem{sonhydro}
D.~T.~Son and A.~O.~Starinets,
Ann. Rev. Nucl. Part. Sci. \textbf{57}, 95-118 (2007)
doi:10.1146/annurev.nucl.57.090506.123120
[arXiv:0704.0240 [hep-th]].
  

\bibitem{mooresound}
  P.~Kovtun, G.~D.~Moore and P.~Romatschke,
  arXiv:1104.1586 [hep-ph].

  \bibitem{kovtun1st}
P.~Kovtun,
JHEP \textbf{10}, 034 (2019)
doi:10.1007/JHEP10(2019)034
[arXiv:1907.08191 [hep-th]].
  
\bibitem{gavassino}
L.~Gavassino, M.~Antonelli and B.~Haskell,
[arXiv:2006.09843 [gr-qc]].

\bibitem{shokri}
M.~Shokri and F.~Taghinavaz,
[arXiv:2002.04719 [hep-th]].


  
\bibitem{ghosts}
G.~Torrieri,
[arXiv:1810.12468 [hep-th]].



\bibitem{reyes}
See the references and introduction of 
  P.~Fries and I.~A.~Reyes,
Phys. Rev. Lett. \textbf{123}, no.21, 211603 (2019)
doi:10.1103/PhysRevLett.123.211603
[arXiv:1905.05768 [hep-th]].

\bibitem{casini}
H.~Casini and M.~Huerta,
Class. Quant. Grav. \textbf{26}, 185005 (2009)
doi:10.1088/0264-9381/26/18/185005
[arXiv:0903.5284 [hep-th]].

\bibitem{kharzeev}
Z.~Tu, D.~E.~Kharzeev and T.~Ullrich,
Phys. Rev. Lett. \textbf{124}, no.6, 062001 (2020)
doi:10.1103/PhysRevLett.124.062001
[arXiv:1904.11974 [hep-ph]].


\bibitem{analogue}
C.~Barcelo, S.~Liberati and M.~Visser,
Living Rev. Rel. \textbf{8}, 12 (2005)
doi:10.12942/lrr-2005-12
[arXiv:gr-qc/0505065 [gr-qc]].


\bibitem{jacobson}
T.~Jacobson,
Phys. Rev. Lett. \textbf{75}, 1260-1263 (1995)
doi:10.1103/PhysRevLett.75.1260
[arXiv:gr-qc/9504004 [gr-qc]].

\bibitem{holo}
G.~Torrieri,
Int. J. Geom. Meth. Mod. Phys. \textbf{12}, no.07, 1550075 (2015)
doi:10.1142/S0219887815500759
[arXiv:1501.00435 [gr-qc]].




\end{thebibliography}
\end{document}